\documentclass[runningheads]{llncs}
\usepackage[utf8]{inputenc}
\usepackage[T1]{fontenc}
\usepackage{amsmath,amssymb}
\usepackage{siunitx}
\usepackage{xspace}
\usepackage[usenames,dvipsnames,table]{xcolor}
\usepackage{graphicx}
\usepackage[style=base,font=small,labelfont=bf]{caption}
\usepackage{subcaption}
\DeclareGraphicsExtensions{.pdf,.png,.eps}
\graphicspath{{figs/}}
\usepackage{tikz}
\usepackage{pgfplots}
\pgfplotsset{compat=1.16}

\usepackage[inline]{enumitem}
\renewcommand{\labelitemi}{$\bullet$}
\usepackage{booktabs}
\usepackage{tabularx}
\usepackage{multirow}
\usepackage{lineno}

\usepackage{mathtools}

\DeclarePairedDelimiter\abs{\lvert}{\rvert}

\def\to{\ensuremath{\rightarrow}} 

\DeclareMathOperator{\dis}{\delta}
\def\Oh{\ensuremath{\mathcal{O}}}
\def\FPILP{\ensuremath{\textup{FP}_{\textup{ilp}}}\xspace}
\def\FAILP{\ensuremath{\textup{FA}_{\textup{ilp}}}\xspace}
\def\FAHEU{\ensuremath{\textup{FA}_{\textup{heu}}}\xspace}
\def\sMM{\textsf{sM-3M}\xspace}
\def\MXL{\textsf{M-3XL}\xspace}
\def\MM{\textsf{M-9M}\xspace}
\def\MS{\textsf{M-18S}\xspace}
\def\CL{\textsf{C-11L}\xspace}
\def\MCL{\textsf{MC-15L}\xspace}
\SetLabelAlign{parright}{\strut\smash{\parbox[t]{\labelwidth}{\raggedleft#1}}}

\usepackage{url}
\usepackage{cite}
\usepackage[hidelinks=true,
			bookmarks=true,
			bookmarksopen=true,
			bookmarksopenlevel=2,
			bookmarksnumbered=true,
			hyperindex=true,
			plainpages=false,
			pdfpagelabels=true,
]{hyperref} 
\usepackage[capitalise,nameinlink]{cleveref}
\newcommand{\etal}{{et~al.}}

\renewcommand{\orcidID}[1]{\href{https://orcid.org/#1}{\includegraphics[scale=.03]{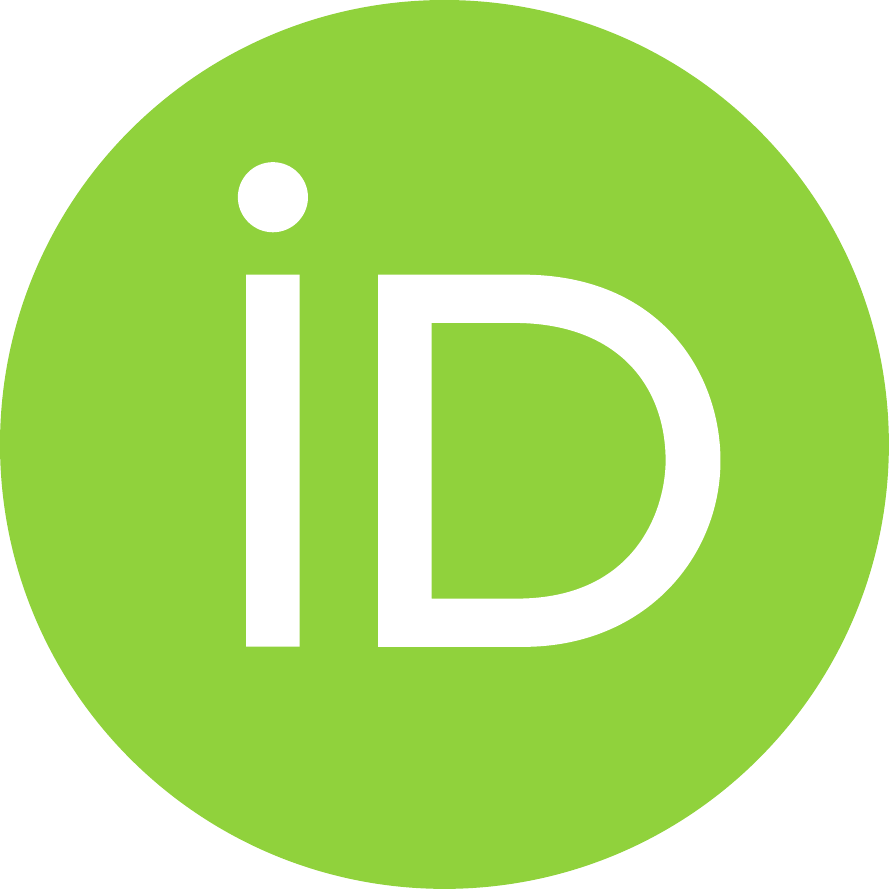}}}

\DeclareTextFontCommand{\bad}{\color{BrickRed}\em}

\usepackage{todonotes}

\title{Algorithms for Floor Planning \texorpdfstring{\\}{} with Proximity Requirements} 

\author{Jonathan~Klawitter \orcidID{0000-0001-8917-5269}
\and Felix~Klesen \orcidID{0000-0003-1136-5673}
\and Alexander~Wolff \orcidID{0000-0001-5872-718X}
}
\authorrunning{Klawitter et al.}
\institute{Universitat Würzburg, Würzburg, Germany}

\begin{document}

\maketitle

\pdfbookmark[1]{Abstract}{Abstract}
\begin{abstract}
  Floor planning is an important and difficult task in architecture.
  When planning office buildings, rooms that belong to the same
  organisational unit should be placed close to each other.  This
  leads to the following NP-hard mathematical optimization problem.
  Given the outline of each floor, a list of room
  sizes, and, for each room, the unit to which it belongs, the aim is
  to compute floor plans such that each room is placed on some floor
  and the total distance of the rooms within each unit is minimized.
  
  The problem can be formulated as an integer linear program (ILP).
  Commercial ILP solvers exist, but due to the difficulty of the
  problem, only small to medium instances can be solved to (near-) optimality.
  For solving larger instances, we propose to split the problem into
  two subproblems; floor assignment and planning single floors.  We
  formulate both subproblems as ILPs and solve realistic problem
  instances.  Our experimental study shows that splitting the problem
  helps to reduce the computation time considerably.
  Where we were able to compute the global optimum,
  the solution cost of the combined approach increased very little.  
  
\end{abstract}
 
\section{Introduction} 

Designing architectural floor plans of an office building 
is a challenging endeavor involving a multitude of tasks.
Among other things, one has to draft a building outline, 
decide on the number of floors, and place rooms, stairs and elevators.
At every step of the process, the planner must meet some 
predefined or implicitly understood requirements.
Therefore, floor planning is a cumbersome process of trial and error
requiring a significant amount of human labour and time.
It is thus of interest to support such manual processes
with (semi-) automated approaches~\cite{LD10}.

We consider a simpler problem.
In our variant of the problem, the planner has already
fixed the outline of each floor, and the position and
width of the corridors.  This is often the case due to fixed lot
sizes as well as distance and construction rules.  We further assume that the
planner has a list of rooms and, for each room, a minimum size.

\begin{figure}[tb]
  \centering
  \includegraphics[page=1]{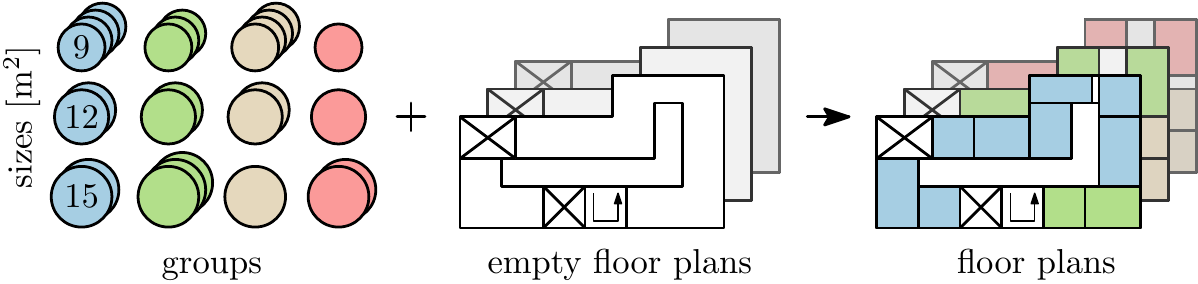}
  \caption{Given a set of rooms, each with its size and group,
  and a set of empty floor plans,
  the \textsc{Floor Planning with Group Proximity} problem
  asks for a placement of the rooms such that
  rooms belonging to the same group are close together.}
  \label{fig:floorPlanning}
\end{figure}

Allen and Fustfeld~\cite{AF75} highlighted the importance
of the architectural layout for communication. 
Based on their observation, some works assume that the planner
has exact information about which pairs of rooms should be placed next
to each other \cite{SPD19}.  
We model proximity relations differently.
In the spirit of Allen and Fustfeld,
we aim to arrange rooms with respect to the
\emph{organisational units} that will use them later.
Technically, the task is to map rooms to location within floors of
the building such that the rooms that belong to the same unit 
or \emph{group} are placed close together; see \cref{fig:floorPlanning}.
We call~this problem \textsc{Floor Planning with Group Proximity}
and define it formally in \cref{sec:model}.

Most variants of floor planning (including ours) are variants
of basic combinatorial \emph{packing problems} such as
\textsc{Knapsack} or \textsc{Bin Packing}; hence they are usually NP-hard.
For this reason it is unlikely that efficient algorithms for floor
planning exist.  Still, algorithms that support architects in this
phase of the planning process are needed since computers are usually
faster than humans in solving NP-hard optimization problems to (near-)
optimality.

\begin{figure}[tb]
  \centering
  \includegraphics{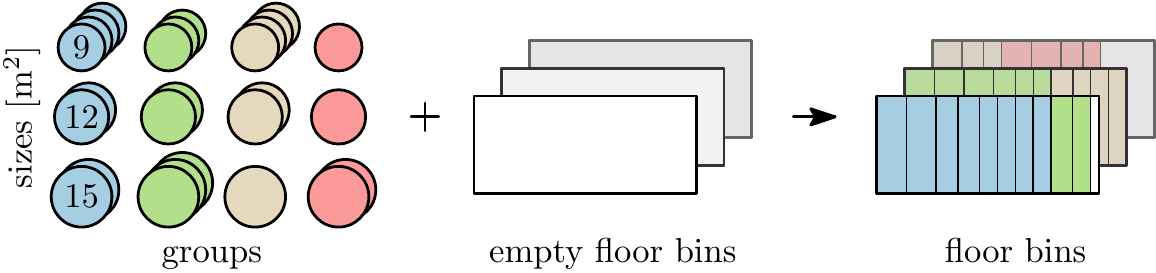}
  \caption{The \textsc{Floor Assignment} problem asks for a mapping
  of rooms to floors such that rooms of the same group 
  appear, preferably, on only few and neighboring floors.}
  \label{fig:floorAssignment}
\end{figure}

We show that \textsc{Floor Planning with Group Proximity}
can be formulated as an integer linear program (ILP).  
Commercial ILP solvers exist, but due to the complexity of our problem, 
only small to medium problem instances can be solved to (near-) optimality.  
For large problem instances, we propose to split
the problem into two independent subproblems. 
First, we map the rooms to floor ``bins'' 
considering the proximity of rooms within their groups;
see \cref{fig:floorAssignment}.
For example, it is preferable to map all rooms of one group to the same floor.
If this is not possible, they should  
be mapped to only few neighboring floors~\cite{AF75}.
We call this problem \textsc{Floor Assignment with Group Proximity}.
Second, we solve \textsc{Floor Planning with Group Proximity} 
for each floor separately.

\paragraph{Contribution.}
We introduce a new type of floor planning problem that occurs when
planning office buildings.  In our model, we assume that 
for each floor we are given its outline, the corridor, and the stairs.
Our objective is to place the rooms into the empty floor plans such that
rooms that belong to the same group are close together.
For the precise problem definition, see~\cref{sec:model}.

We formulate our floor planning problem as an ILP;
see \cref{sec:ilp:fp}.  According to our computational
experiments, it takes too long to find optimal solutions of this
global formulation for large instances of the problem.  Thus we split
the global problem into two independent subproblems; floor assignment
and (single) floor planning, which we then solve by separate ILPs (see
\cref{sec:ilp:fa}).  While optimal solutions for the subproblems do not
necessarily lead to optimal solutions of the global problem, our
computational experiments show that the loss in the global objective
is acceptable and the runtime improvements are considerable.  We also
test a simple greedy heuristic for the floor assignment subproblem;
see \cref{sec:experiments}.

While the runtime of ILP solvers is difficult to predict, linear
optimization is quite powerful and allows the user to add additional
constraints (such as ``these two groups must go to the basement'')
easily.  If a user is dissatisfied with a solution that is provably
optimal, then it is not the algorithm to be blamed~-- but the model.
In other words, the objective must be changed or further constraints
must be added in order to exclude solutions with certain undesired
features.  Once the model has been settled, it is possible to drop the
requirement of optimal solutions; quite often ILP solvers quickly find
near-optimal or even optimal solutions, but then need very long to
\emph{prove} their optimality; see \cref{tbl:fa:runtimes}.

We think that our work has the potential to support architects in
finding sustainable solutions both for themselves and the users of
their designs.

\section{Related Work}

The literature on floor planning problems is quite diverse.
On the one hand, a panoply of different algorithmic methods 
have been used to tackle such problems.
They range from logic programming~\cite{Kov91}, 
constraint programming~\cite{Cha94}, 
quadratic programming~\cite{MCP02}, 
over using shape rules and grammars\cite{Dua05,HK14,VCS18} 
or graph-theoretic tools~\cite{MM10,SD17,SPD19} 
to evolutionary algorithms~\cite{kk-gfplk-GA10,HK14} to name just a few.
We refer the interested reader to the surveys by Del Río-Cidoncha~\etal~\cite{DRC+07}
and by Lobons and Donath~\cite{LD10}.  
On the other hand, nearly every paper considers a different problem definition.
We can group these variants as being purely combinatorial (where
only room sizes matter), more geometric (where the actual
shapes and their aspect ratios matter), or purely topological
(where only adjacencies matter).  
Our problem variant includes both geometric and,
with the group proximity constraints, topological aspects.

A purely combinatorial version of floor planning is the well-known
\textsc{Bin Packing} problem, where items of different sizes
must be packed into bins, each of a fixed capacity,
in a way that minimizes the number of bins used.
Bergman \etal~\cite{BCM19} consider a variant called
\textsc{Bin Packing with Minimum Color Fragmentation},
where each item is associated with a color.
Then the goal is to find a bin packing where
items of a common color are placed in the fewest number of bins possible.
This problem is closely related to \textsc{Floor Assignment with Group Proximity}
though the quality of a solution is measured differently. 
If rooms are set to be rectangular and their sizes are
prescribed (with aspect ratios of rooms either fixed or bounded),
we are in the range of \textsc{Rectangle Packing} problems~\cite{Bor13,HK13}.

Marson and Musse~\cite{MM10} showed how to generate floor plans for
residential houses where just a few adjacencies are specified.
They prescribed room sizes and
used squarified treemaps to subdivide the fixed building outline.
Knecht and König~\cite{kk-gfplk-GA10} used an evolutionary approach to
generate subdivisions of a given rectangle into a given number 
of smaller rectangles (rooms).  
In a second step, they used a genetic algorithm to
change the topology of the resulting subdivision and to obtain the
desired adjacencies. 
More recently, Shi et al.~\cite{SSHW20} used a
Monte-Carlo tree search to evaluate and select promising candidates
among many floor plans that they build room by room.
Due to their runtime, some of these approaches only work for
small houses, but not for large office buildings.

If all allowed room adjacencies are already prescribed,
the input of a floor planning problem takes 
the form of a (well-behaved) triangulated embedded planar graph.
The corresponding floor plan is then called a
\emph{rectangular dual}~\cite{EMSV12}.
Formally, a rectangular dual is
a dissection of a rectangle into smaller rectangles such that the
adjacency graph of the smaller rectangles is the given embedded graph.
Upasani et al.~\cite{USS20} presented an iterative
procedure that takes a rectangular dual and lower and upper bounds on
the room dimensions (in x- and y-direction) as input.  
The algorithm then optimizes the layout by alternatingly computing network flows in
the graphs that represent the horizontal and the vertical contacts
between the rectangular rooms.  
%
Instead of expecting an adjacency matrix as input, Simon~\cite{Sim17}
generated floor plans with a genetic algorithm that minimizes traffic
flow between rooms, e.g., class rooms in schools.

A problem related to floor planning is the \emph{facility layout}
problem where facilities have to be arranged efficiently within an
organization. In contrast to floor planning, facility layout is less
about subdividing a given building, but about the placement of the
facilities and the resulting paths between them. The aim is to place
facilities such that the paths allow for low material handling costs,
short lead time, and high productivity.  
We refer interested readers to the surveys by Meller and Gau~\cite{MG96} 
and by Drira \etal~\cite{DPHG06}.  
Like for floor planning, there also exist multi-floor variants 
where departments have to be placed on floors and
convenient positions for lifts have to be found; 
see a recent survey by Ahmadi \etal~\cite{APJ17}.  
Interestingly, Meller and Bozer~\cite{MB97} also suggest a two-stage 
approach for the multi-floor facility layout problem.

Barth et al.~\cite{bfklnopsuw-swcrh-LATIN14} and Bekos et
al.~\cite{bdfkkpsw-iaabc-Algorithmica17} considered a related
problem that was motivated by drawing semantic word clouds.  Given a
set of rectangular shapes with fixed sizes and an adjacency graph
defined on these shapes, place the shapes such that no two shapes
overlap. Their objective was to realize the maximum number of desired
adjacencies as side contacts of the rectangular shapes.

\section{Our Model and Problem Definitions} 
\label{sec:model}
In this section, we describe a model for floor plans, 
allowed room placement, and group proximity 
that is tailored to our ILP approach. 
We then make the problem statements precise.

\begin{figure}[tb]
  \centering
  \includegraphics{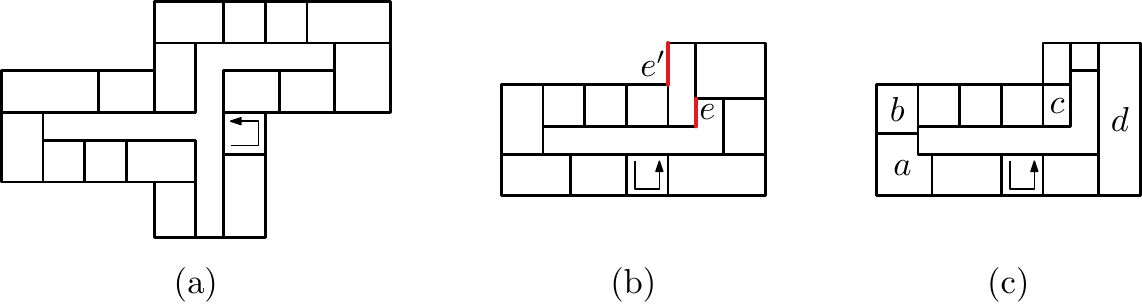}
  \caption{Three orthogonal floor plans: (a)~is valid; (b)~is invalid
    since the corridor wall~$e$ does not overlap with the outer 
    wall~$e'$ (when projected to the y-axis);
    and (c)~is invalid since room~$a$ is not rectangular,
    room~$b$ shares not enough wall with the corridor for a door,
    room~$c$ has no window, and room~$d$ occupies two building corners.}
  \label{fig:basicFloorplan}
\end{figure}

\paragraph{Floor plans.}
A \emph{floor plan} (of a single floor)
describes a subdivision of the floor's \emph{outline}, a simple polygon,
into smaller polygons by inner walls.
Each of the smaller polygons represents a room, the corridor, stairs, etc.
Here we consider only \emph{orthogonal floor plans}, 
that is, each wall is drawn either horizontally or vertically.
We require that the \emph{corridor} runs parallel to the outline, 
that is, the polygon describing the corridor is combinatorially the same as the polygon describing the outline; 
see \cref{fig:basicFloorplan}(a).
In particular, to keep our model simple, each vertical or horizontal segment of the corridor must overlap vertically or horizontally, respectively, 
with its combinatorial counterpart of the outline; see \cref{fig:basicFloorplan}(b) for a counterexample.
For the following requirements, see \cref{fig:basicFloorplan}(c).
We insist that all \emph{rooms} are rectangular
and stretch from the corridor to the outline.
To ensure that each room gets a door and a window,
we require that the room has a certain minimum overlap
with the corridor and with the outline.
To keep the model simple
a room may occupy at most one corner 
(but see \cref{sec:conclusion}).

\paragraph{Empty floor plans.}
Part of the input of the floor planning problem are \emph{empty floor plans}.
Geometrically, we require that an empty floor plan consists of the outline, the corridor, 
at least one room that represents \emph{stairs} or elevators, 
and potentially other \emph{blocked areas}.
Stairs connect the individual floors of a building with each other.
Blocked areas represent, for example, sanitary facilities, 
which are often located at the same place on each floor.
The \emph{capacity} of an empty floor plan is the size 
of the area not covered by the corridor and blocked rooms. 

We model an empty floor plan as follows.
We subdivide the unoccupied area between the outline and the corridor 
into rectangles of two types; see \cref{fig:emptyFloorPlans}.
Each pair that consists of a corner of the outline and the corresponding corner of the corridor
spans a \emph{corner rectangle}. 
The rectangles that form the remaining area are
\emph{edge rectangles}.
For the sake of brevity, we call corner and edge rectangles
simply \emph{corners} and \emph{edges}, respectively.
For each corner $v$, its capacity $\kappa_v$ is its area.
Similarly, for each edge $e$, the capacity of $e$ is denoted by $\kappa_e$.
The sum of these capacities is the capacity of the empty floor plan.

\paragraph{Room placement.}
We now describe a set of rules that define how rooms 
can be placed into empty floor plans to obtain valid floor plans. 
Recall that each room comes with a (minimum) \emph{size}.
A \emph{room placement} is a mapping of rooms into the unoccupied area,
defined by the corners and edges, and their capacities.
A valid room placement satisfies the following conditions. 
\begin{enumerate}[label=(P\arabic*),leftmargin=*]
  \item \label{cs:p1} 
  Each room is either mapped to an edge 
or to a pair consisting of a corner and an adjacent edge.
  \item \label{cs:p2} 
  For each corner $v$, at most one room may be mapped to $v$
  and this room must occupy $v$ fully.
  Moreover, the size of the room must exceed the capacity of $v$. 
  This excess must be enough for the room to admit a door and a window.
  \item \label{cs:p3} 
  For each edge $e$ with adjacent corners $v$ and $v'$, 
  the sizes of rooms mapped to $e$ plus the total excess of the rooms
  mapped to $(e,v)$ and $(e,v')$ may not exceed the capacity of~$e$.
\end{enumerate}
Note that \cref{cs:p3} implies that a room may occupy at most one corner.
Further restrictions on room placements are possible,
for example, we could forbid rooms to be placed at an edge 
where its aspect ratio becomes undesirably large.

We want to point out that a room placement does not
fix the positions of rooms along an edge,
but only allocates the necessary space.
Hence, the exact positions of such rooms 
need to be computed in a post-processing step.
We return to this matter in \cref{sec:experiments}.

\paragraph{Group proximity.}
Given a room placement, we describe how to measure the proximity 
of the rooms within a group, for short, the \emph{group proximity}.
To this end, let $V$ be the set of corners
and $E$ the set of edges.
Let $O = V \cup E$ be the set of \emph{objects} (that is, vertices and edges).
For two objects $o, o' \in O$, 
let~$\dis_{o,o'}$ denote the distance of~$o$ and~$o'$.
For example, $\dis_{o,o'}$ could be the length of a path from $o$ to $o'$
along the corridor (possibly using stairs).
In general, however, the planner can set the distances as they see fit
provided that, for~$o = o'$, $\dis_{o,o'} = 0$.
For a group $g$, we then define the proximity of~$g$
as the sum of distances~$\dis_{o,o'}$ over all pairs~$o, o' \in O$,
where both~$o$ and~$o'$ contain at least one room of~$g$.

\paragraph{Problem definitions.}
We now define our two problems.
To this end, let~$G$ be the set of groups, and let~$S$ be the set of room sizes.
For each~$g \in G$ and each~$s \in S$, 
let~$\rho_{g,s}$ denote the number of rooms of size $s$ that belong to group $g$. 

First, we define the problem \textsc{Floor Planning with Group Proximity}.
Let~$V$ and~$E$ be the sets of corners and edges, respectively,
that together with their distance matrix~$\dis$ and capacities~$\kappa$
describe the available empty floor plans.
A feasible solution for the problem is a valid placement 
of all rooms into~$V$ and~$E$ with respect to~$\kappa$, as defined above.
We say that a feasible solution is optimal if it minimizes the
sum of proximities over all groups in~$G$ with respect to~$\dis$.

Next, we define the problem \textsc{Floor Assignment with Group Proximity}.
Let~$F$ be the set of available floors.
Each floor $f$ in $F$ has a capacity~$\kappa_f$, and
for two floors $f, f'$ in $F$, $\dis_{f,f'}$ denotes their distance.
For example, if all floors belong to a single building,
the $i$th floor and $j$th floor could have distance $\abs{j - i}$.
A feasible solution for the problem is an assignment of all rooms
to the floors in~$F$ such that no floor~$f \in F$ is overfilled with respect to~$\kappa_f$.
Given a feasible solution and a group~$g \in G$, 
let~$F_g \subseteq F$ be the set of floors that contain a room of group~$g$.
We say that a feasible solution is optimal if it minimizes~$\sum_{g \in G}\sum_{f,f' \in F_g} \dis_{f,f'}$.

For ease of reading, we refer to these two problems from now on
simply with \textsc{Floor Planning} and \textsc{Floor Assignment}.

\section{An ILP for Floor Planning}  
\label{sec:ilp:fp}
Linear programming is a popular tool to solve combinatorial optimization problems.
A \emph{linear program (LP)} consists of (i) real-valued variables $x_1, \dots, x_n$, 
(ii)~a target function that is restricted to be linear in the variables
(e.g., minimize $c_1x_1 + \dots + c_nx_n$ for some constants $c_1, \dots, c_n$), 
and (iii) a set of linear constraints ($a_{i,1}x_1 + \dots + a_{i,n}x_n \ge b_i$ for $i = 1, \dots, m$,). 
Linear programs can be solved efficiently \cite{k-nptalp-Comb84}.
A \emph{(mixed-)integer linear program (ILP)} is a generalization of a linear
program where some variables can be restricted to integer values.
In particular, binary ``decision'' (that is, 0--1) variables can be used.
This makes it possible to encode NP-hard combinatorial optimization problems.
Consequently, ILPs cannot be solved efficiently in general.
In practice, however, small and medium-sized instances 
of such problems can often be solved relatively fast~\cite{sz-liotp-15}.
For example, we can solve the below ILP formulation for \textsc{Floor Planning} for a single floor
with $40$ rooms and three groups in under one second.

We now describe how to formulate \textsc{Floor Planning} as an ILP.
The input of the ILP consists of the empty floor plans
given by the sets~$V$ and~$E$ of corners and edges, respectively,
their adjacency relations,
their distances $\delta$, and their capacities~$\kappa$.
Let $O = V \cup E$ be the set of
\emph{objects}, that is, the corners and edges.
The ILP further gets the set~$G$ of groups, the set~$S$ of room sizes,
and the room quantities~$\rho$ as input.
Note that the ILP views all numbers in the input as constants.  
(Since the distances are part of the input,
they hide the number of floors from the ILP. 
The distances are also not restrained to the actual geometry 
of the floor plans and can thus be set as desired by the planners.)

We need the following variables, all of which are binary except for
the first one, which is an integer.  To help intuition, we specify the
intended meaning of the variables.
\begin{align*}
  x_{g,s,e} \geq 0 & \text{ denotes the number of group-$g$, size-$s$ 
                     rooms placed at edge $e$.} \\[-.3ex] 
  y_{g,s,v,e} = 1  &\Leftrightarrow \text{a group-$g$, size-$s$ room occupies corner $v$ and extends into edge $e$.}  \\[-.3ex]
  z_{g,o} = 1      &\Leftrightarrow \text{a room of group $g$ is at object $o$
                     (which is an edge or a vertex).} \\[-.3ex]
  u_{g,o,o'} = 1   &\Leftrightarrow \text{rooms of group $g$ are
                     at objects~$o$ and~$o'$.}
\end{align*}
If placing a room of size~$s$ into corner~$v$ along
edge~$e$ would not allow this room to have a door or a window, 
we set $y_{g,s,v,e} = 0$ for every group~$g$.
(This partially enforces room placement condition \ref{cs:p2}.)
Similarly, if placing a room of size~$s$ at edge~$e$ 
would make the room's aspect ratio too extreme, 
we could set $x_{g,s,e} = 0$ for every group~$g$.

Recall that the problem asks to minimize the sum of distances between 
pairs~$o, o'$ of objects that contain rooms from the same group $g$.
The triples $g, o, o'$ that contribute to this sum
are those with $u_{g,o,o'} = 1$.
Hence, the objective function of our ILP for
\textsc{Floor Planning} (\FPILP) is:
\begin{align*}
  \text{minimize} & \hspace*{-1ex}\sum\limits_{g \in G, o, o' \in O} \hspace*{-1ex} u_{g,o,o'} \cdot \dis_{o,o'},
\end{align*}
which is subject to the following constraints. 
The first three constraints enforce the room placement conditions \ref{cs:p1} to \ref{cs:p3}.
\begin{align*}
  \intertext{\labelitemi$\,$ Place all rooms \ref{cs:p1}:\vspace*{-1ex}}
  \sum\limits_{e \in E} x_{g,s,e} + \hspace*{-1.5ex}
  \sum\limits_{\substack{v \in V\\[.3ex] e \text{ adjacent to } v}}\hspace*{-2.5ex} y_{g,s,v,e} &= \rho_{g,s}       && \text{for } g \in G, s \in S \\
  \intertext{\labelitemi$\,$ Place at most one room in a corner $v$ \ref{cs:p2}:\vspace*{-1ex}}
  \sum\limits_{\substack{g \in G, s \in S,\\[.3ex] e \text{ adjacent to } v}}\hspace*{-2ex} y_{g,s,v,e}&\leq 1 && \text{for } v \in V \\
  \intertext{\labelitemi$\,$ Do not overfill an edge $e$ \ref{cs:p2}--\ref{cs:p3}; that is,
  the sum of the sizes of all rooms fully placed at $e$ and those
  extending into $e$ from a corner must not exceed~$\kappa_e$:
  \vspace*{-1ex}} 
  \sum\limits_{g \in G, s \in S}\hspace*{-1.5ex} x_{g,s,e} \cdot s + \hspace*{-1.5ex}
  \sum\limits_{\substack{g \in G, s \in S,\\[.3ex] v \text{ adjacent to } e}}\hspace*{-2.5ex}
  y_{g,s,v,e} \cdot (s - \kappa_v) &\leq \kappa_e 									&& \text{for } e \in E \\
  \intertext{Note that the room sizes in $S$ are constants from the
  point of view of the ILP.  We need the following constraints to set the
  binary variables of types~$u$ and~$z$.}
  \\[-5ex]
  \intertext{\labelitemi$\,$ Force $z_{g,e}$ to 1 if a room of group~$g$ is placed at edge~$e$:\vspace*{-1ex}}
  x_{g,s,e}/\rho_{g,s} &\le z_{g,e} 	&& \text{for } e \in E, g \in G, s \in S \\[-1ex]
  \intertext{\labelitemi$\,$ Force $z_{g,v}$ to 1 if a room of group~$g$ is placed at corner~$v$:\vspace*{-1ex}}
  y_{g,s,e,v} &\le z_{g,v} && \text{for } v \in V, g \in G, s \in S, \\[-1ex]
  &&& e \text{~incident to~} v \\[-1ex]
  \intertext{\labelitemi$\,$ Force $u_{g,o,o'}$ to 1 if rooms of group $g$ are placed at objects~$o$ and~$o'$:\vspace*{-1ex}}
  z_{g,o} + z_{g,o'} - 1 &\le u_{g,o,o'} && \text{for } o, o' \in O, g \in G
\end{align*}
Note that we define only lower bounds for variables of types $u$ and $z$ here.
However, setting, for example, $u_{g,o,o'}$ to~1 
\textit{without} rooms of group~$g$ being placed at both
objects~$o$ and~$o'$ would increase our objective function.  
Hence, the ILP solver sets $u_{g,o,o'}$ to~0 in this case.

\paragraph{Dealing with unsolvable instances.}
Consider an instance of \textsc{Floor Planning}
where the empty floor plans have a total capacity $K$ 
and all rooms together require an area of $A$.
If the rooms require more area than available (that is, $A > K$), 
then clearly, there is no solution.
However, even if enough area is available (that is, $A \leq K$)
there may not necessarily exist a solution,
because the rooms placed at the same edge $e$ 
may not fill up $e$ completely and 
hence available area remains unused. 
Moreover, there might not be enough large rooms too occupy all corners.
In general, this is more likely to happen when 
there is not much spare area available.
One way to deal with such unsolvable instances is to scale
down room sizes, which effectively decreases~$A$.

\section{An ILP for Floor Assignment}  
\label{sec:ilp:fa}
Due to the complexity of the \textsc{Floor Planning} problem,
our ILP formulation from the previous section 
can only be solved for instances of moderate size.
We thus propose to split large instances into smaller ones.
More precisely, we solve the respective \textsc{Floor Assignment} problem
that splits a large multi-floor instance into individual floors.
As a result, we get single-floor instances of \textsc{Floor Planning} 
that can usually be solved within a reasonable amount of time.

We now describe an ILP for the \textsc{Floor Assignment} problem.
Recall that the problem asks us to assign every room to one of the floors 
such that the rooms of each group are assigned only to few floors that are close together.
The input is given by the set $F$ of floors, 
their distances~$\delta$, their capacities~$\kappa$,
the set~$G$ of groups, the set~$S$ of room sizes,
and the room quantities~$\rho$.
Note that distances of floors set in $\dis$ do not need to grow linearly.
In particular, one may set distances to model that people take the stairs to go up one or two floors,
but take the elevator for more floors. 

We need the following variables all of which are binary except for
the first one, which is an integer.
\begin{align*}
  x_{g,s,f} \geq 0 & \text{ denotes the number of group-$g$, size-$s$ rooms assigned to floor $f$.} \\[-.3ex]
  z_{g,f} = 1 &\Leftrightarrow \text{a room of group $g$ is assigned to floor $f$.} \\[-.3ex]
  u_{g,f,f'} = 1 &\Leftrightarrow \text{rooms of group $g$ are assigned to floors~$f$ and~$f'$.}
\end{align*}
Then our ILP for \textsc{floor assignment} (\FAILP) is as follows.
\begin{align*}
  \text{Minimize} & \sum\limits_{g \in G, f, f' \in F}\hspace*{-1.5ex} u_{g,f,f'} \cdot \dis_{f,f'} 
\end{align*}
subject to  the following constraints.
\vspace*{-2ex}
\begin{align*}
  \intertext{\labelitemi$\,$ Assign all rooms: \vspace*{-2ex}}
  \sum\limits_{f \in F} x_{g,s,f} 						& = \rho_{g,s}					&& \text{for } g \in G, s \in S \\[-2ex]
  \intertext{\labelitemi$\,$ Do not overfill any floor: \vspace*{-2ex}}
  \sum\limits_{g \in G, s \in S}\hspace*{-2ex} x_{g,s,f}\cdot s 	& \leq \kappa_{f} 	&& \text{for } f \in F \\[-2ex]
  \intertext{\labelitemi$\,$ Force $z_{g,f}$ to 1 if a room of group $g$ is assigned to floor $f$: \vspace*{-2ex}}
  x_{g,s,f}/\rho_{g,s} &\le z_{g,f} && \text{for } f \in F, g \in G, s \in S \\[-2ex]
  \intertext{\labelitemi$\,$ Force $u_{g,f,f'}$ to 1 if rooms of group $g$ are assigned to floor $f$ and $f'$: \vspace*{-2ex}}
  z_{g,f} + z_{g,f'} - 1 & \le u_{g,f,f'} && \text{for } f, f' \in F, g \in G
\end{align*}

\section{A Heuristic for Floor Assignment} 
\label{sec:heuristic}
In this section, we propose a heuristic  for the \textsc{Floor Assignment} problem.
Our heuristic, which we call \FAHEU, tries to distribute the rooms evenly among the floors.
This is motivated by the observation 
that an instance of \textsc{Floor Planning} with a nearly full floor
is less likely to have a solution.
Roughly speaking, \FAHEU consists of three steps, namely,
(i) reserving area on each floor, 
(ii) allocating space to the groups, 
and (iii) distributing the rooms of each group to the corresponding allocated space.
We explain these steps now more precisely.

The first step of \FAHEU works as follows.
Recall that $F$ is the set of floors. 
Now let $K$ be the sum of the capacities of all floors,
let $A$ be the sum of sizes of all rooms,
and thus $K-A$ is the excess area of the building.
Then \FAHEU reserves on each floor an area of size $(K - A)/ \abs{F}$,
that is, an equal proportion of the excess area.

\begin{figure}[htb]
  \centering
  \includegraphics{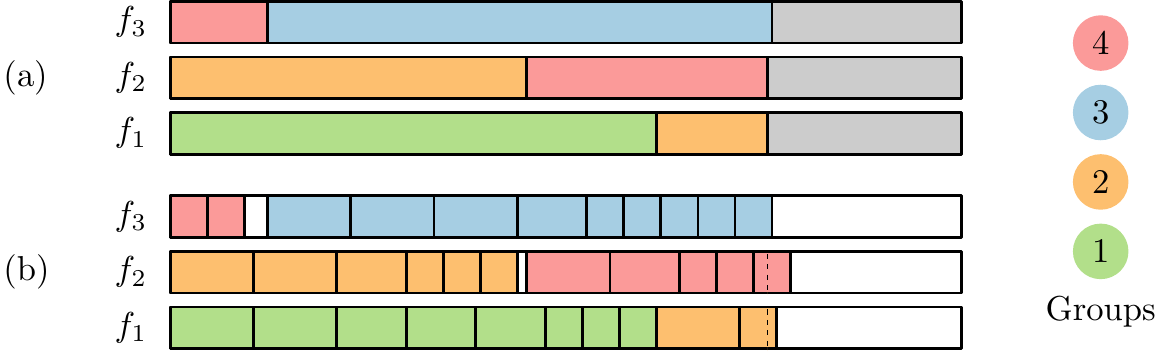}
  \caption{\FAHEU applied to a small problem instance (\sMM, see \cref{sec:testdata}).
  (a) Free space (grey) is evenly distributed to the three floors 
  and the four groups get area allocated one after the other.
  (b) Allocated space is converted into rooms whereat at most one room per floor uses reserved space.}
  \label{fig:fa:heuristic}
\end{figure}

In the second step, the remaining area is allocated to the groups.
To this end, \FAHEU iterates through groups and floors concurrently.
More precisely, suppose that currently group~$g$ and floor~$f$ are handled.
Then as much area of~$f$ is allocated to~$g$
as either~$f$ has available or as~$g$ still requires.
Accordingly, \FAHEU either proceeds with the next floor or the next group.
For an example, consider \cref{fig:fa:heuristic}(a), 
where group~1 requires \SI{105}{\square\meter} of the unreserved \SI{129}{\square\meter} on floor~$f_1$.
Proceeding with group~2, which requires \SI{101}{\square\meter},
the heuristic first allocates the remaining \SI{24}{\square\meter} of~$f_1$ to group~2
and then continues with floor~$f_2$.

In the last step, \FAHEU converts the allocated areas of the second step
into an assignment of rooms to floors as follows.
If the whole allocated area of a group~$g$ is on one floor, 
then the area can be straightforwardly partitioned into the rooms of~$g$
(like for group~1 in \cref{fig:fa:heuristic}(a--b)).
Otherwise, if~$g$ is the last group that got area allocated on a floor~$f$
(like group~2 on floor~$f_1$ in the example),
then \FAHEU repeatedly assigns the largest remaining (that is, unplaced) room of~$g$ on floor~$f$
that fits inside the allocated area.
If this is no longer possible, but the allocated area of $g$ on~$f$ is not fully used up yet,
\FAHEU assigns the smallest remaining room of $g$ to~$f$.
In the example, first the largest room and then the smallest room of group~2 are assigned to~$f_1$.
In this way, \FAHEU processes floor after floor and group after group.

Note that as long as the reserved area on each floor is at least as large as the largest room size,
\FAHEU always returns a valid solution.
We always place the smallest room at the end of a floor in order to equally distribute the excess area.

The first step takes $\Oh(\abs{G} \cdot \abs{S} + \abs{F})$ time
since we need to sum the sizes of all rooms and the capacities of all floors.
For the second step, we need $\Oh(\abs{G} + \abs{F})$ time to greedily allocate the area of the floors to groups.
The third step takes $\Oh((\abs{G} + \abs{F}) \cdot \abs{S})$ time
since every room is assigned successfully once and each size may be tried unsuccessfully once per floor.
Therefore, the total runtime of \FAHEU is also $\Oh((\abs{G} + \abs{F}) \cdot \abs{S})$.

\section{Test Data} 
\label{sec:testdata}
To get a rough idea of realistic problem instances,
we considered the situation at two institutes 
at the University of Würzburg; Mathematics and Computer Science.
The actual floor plans of the mathematics buildings inspired the room sizes of our instances.
We mapped each type of staff, such as professors and research assistants, to a different room size.
Accordingly, we modeled the institutes' chairs as groups, with their respective staff. 
Based on this data, we built problem instances
that vary in the number of groups, 
in the number of floors, in the floor sizes,
and in the number of different room sizes. 

\begin{figure}[htb]
  \centering
  \begin{subfigure}[b]{.43 \linewidth}
		\centering
		\includegraphics[page=1]{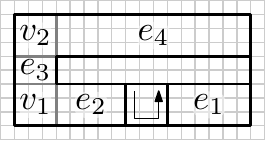}
		\caption{Floor plan $f_\mathrm{S}$  has capacity \SI{99}{\square\meter}.}
		\label{fig:emptyFloorPlans:I}
	\end{subfigure}
	\hfill
	\begin{subfigure}[b]{.55 \linewidth}
		\centering
		\includegraphics[page=2]{emptyFloorPlans}
		\caption{Floor plan $f_\mathrm{M}$  has capacity \SI{171}{\square\meter}.}
		\label{fig:emptyFloorPlans:II}
	\end{subfigure}
	\hfill
	
	\smallskip
	
	\begin{subfigure}[b]{.43 \linewidth}
		\centering
		\includegraphics[page=3]{emptyFloorPlans}
		\caption{Floor plan $f_\mathrm{L}$  has capacity \SI{318}{\square\meter}.}
		\label{fig:emptyFloorPlans:III}
	\end{subfigure}
	\hfill
	\begin{subfigure}[b]{.55 \linewidth}
		\centering
		\includegraphics[page=4]{emptyFloorPlans}
		\caption{Floor plan $f_\mathrm{XL}$ has capacity \SI{512}{\square\meter}.}
		\label{fig:emptyFloorPlans:IV}
	\end{subfigure}
  \caption{The four empty floor plans used in our problem instances.}
  \label{fig:emptyFloorPlans} 
\end{figure}

We first treat floor plans.
We designed four different empty floor plans, 
$f_\mathrm{S}$, $f_\mathrm{M}$, $f_\mathrm{L}$, $f_\mathrm{XL}$ in order of increasing size; 
see \cref{fig:emptyFloorPlans}.
As an example, \cref{tbl:floordistancematrix} shows the distance matrix of $f_\mathrm{S}$.
For all instances, we set the distance of the $i$th and the $j$th
floor to $\abs{j-i}\cdot\SI{20}{\meter}$ \label{page:distance}
(seen as a penalty)
and calculate the distances of edges and vertices of different floors with their distances to the stairs.

Note that the floor capacities correlate with the floor complexities
in terms of numbers of edges and corners.
We want to point out that \FPILP does not take the 
actual geometry of a floor into account but only its complexity, the capacities,
and the distance matrix.
In particular, our problem definition does not insist that 
the same empty floor plan is used for every floor.
For simplicity and comparability of the results, in each of our test instances, we do use the same empty floor plan for every floor.

We define the following sets of groups.
\begin{description}[leftmargin=!,labelwidth=7.75mm]
\item[\textsf{M}:] A set of groups based on the chairs of the Institute of Mathematics. 
It contains 11 groups with a total of 122 rooms of sizes 8, 15, or \SI{18}{\square\meter}. The exact groups are shown in \cref{tbl:groups}.
\item[\textsf{C}:] A set of groups based on the chairs of the Institute of Computer Science
that contains 9 groups with a total of 177 rooms of sizes 10, 15, 20, and \SI{25}{\square\meter}; see again \cref{tbl:groups}.
\item[\textsf{sM}:] A small set of groups that contains the math groups 1, 2, 4, and 11.
\item[\textsf{MC}:] A large set of groups; the union of the sets \textsf{M} and \textsf{C}.
It thus contains 20 groups with  a total of 299 rooms of six different sizes.
\end{description}

\begin{table}[htb]
  \parbox[b]{.46\linewidth}{%
  \centering
  \setlength{\tabcolsep}{1ex}
  \begin{tabular}{@{}lr@{~~~~}rrrr@{~~~~}rl@{}}
    \toprule
    					&& $e_1$ & $e_2$ & $e_3$ & $e_4$ & $v_1$ & $v_2$ \\ 
    \midrule
    \parbox[t]{9mm}{\multirow{4}{*}{Edges}} 
                     & $e_1$ & 0 & 5 & 7 & 2 & 8 & 8 \\
					 & $e_2$ & 5 & 0 & 1 & 2 & 0 & 2 \\
					 & $e_3$ & 7 & 1 & 0 & 1 & 0 & 0 \\
					 & $e_4$ & 2 & 2 & 1 & 0 & 2 & 0 \\ 
    \midrule
    \parbox[t]{11mm}{\multirow{2}{*}{Corners}} 
                     & $v_1$ & 8 & 0 & 0 & 2 & 0 & 1 \\
					 & $v_2$ & 8 & 2 & 0 & 0 & 1 & 0 \\
    \midrule
			Stairs   && 2 & 2 & 5 & 2 & 5 & 5 \\
    \bottomrule
  \end{tabular}
  \medskip
  \caption{Distances in the empty floor plan  $f_\mathrm{S}$.}
  \label{tbl:floordistancematrix}
  }
  \hfill
  \parbox[b]{.5\linewidth}{
  \centering
  \setlength{\tabcolsep}{1ex}
  \begin{tabular}{@{}rr@{~~~~}rrr@{~~~~}rrrr@{}}
    \toprule
   	\multicolumn{2}{@{}l}{Institute} & \multicolumn{3}{c}{Math} & \multicolumn{4}{c}{CS} \\
    \multicolumn{2}{@{}l}{Sizes [\SI{}{\square\meter}]} & 8 & 15 & 18 & 10 & 15 & 20 & 25\\\midrule
	\parbox[t]{10mm}{\multirow[c]{11}{*}{\hspace{4mm} \rotatebox[origin=r]{90}{Groups}}}
					&  1 &  3 &  3 &  2 &  6 &  8 &  2 &  1\\
					&  2 &  4 &  1 &  3 &  2 & 19 &  2 &  2\\
					&  3 &  5 &  1 &  3 &  3 & 10 &  1 &  1\\
					&  4 &  5 &  1 &  1 &  0 &  3 &  1 &  1\\
					&  5 &  9 &  1 &  2 &  2 & 14 &  3 &  2\\
					&  6 & 11 &  1 &  5 &  3 & 17 &  2 &  2\\
					&  7 & 16 &  1 &  3 &  0 & 15 &  2 &  2\\
					&  8 &  8 &  1 &  3 &  3 & 28 &  1 &  2\\
					&  9 &  4 &  1 &  3 &  1 & 14 &  1 &  1\\
					& 10 &  7 &  1 &  4 \\
					& 11 &  8 &  1 &  3 \\
    \bottomrule
  \end{tabular}
  \medskip
  \caption{The mathematics and computer science groups.}
  \label{tbl:groups}
}
\end{table}

Finally, we can define our six problem instances.  We have three
medium-sized instances that use the set \textsf{M} of math groups 
but floors of different sizes.  Furthermore, we have a small, a large,
and a very large instance.

\begin{description}[leftmargin=!,labelwidth=15.75mm]
  \item[\MS:] Math groups \textsf{M} and 18 copies of
  $f_\mathrm{S}$.
  \item[\MM:] Math groups \textsf{M} and 9 copies
  of~$f_\mathrm{M}$.  The problem instance is closest to the actual
  situation at the Institute of Mathematics.
  \item[\MXL:] Math groups \textsf{M} and 3 copies
  of~$f_\mathrm{XL}$.
  \item[\sMM:] Subset of math groups \textsf{sM} and 3 copies
  of~$f_\mathrm{M}$.
  \item[\CL:] Computer science groups \textsf{C} and 11 copies
  of~$f_\mathrm{L}$.  This instance is larger and more complex than
  the math instances as it has more rooms and different room sizes.
  \item[\MCL:] Combined groups \textsf{MC} and
  15 copies of~$f_\mathrm{L}$.
\end{description}

\section{Experimental Evaluation} 
\label{sec:experiments} 
In this section, we evaluate our three approaches for the \textsc{Floor Planning} problem
on our problem instances from \cref{sec:testdata}.
First, we examine the performance of \FPILP on its own.
Then we combine \FPILP with either \FAILP or \FAHEU,
and discuss the strengths and limitations of these approaches. 
Finally, we detail how we post-process \FPILP solutions.

\paragraph{Using an ILP for floor planning.}
All ILP tests were run using CPLEX with OPL on a virtual machine that
runs Ubuntu 18.04 with 16 cores and 96\,GB of RAM.
After preliminary experiments, we decided to use settings of CPLEX that resemble a depth-first search
with a focus on finding feasible solutions.
We tested \FPILP twice on each of the six problem instances,
once with a realistic time limit of one hour 
and once with a time limit of \SI{12}{\hour} for comparison. 
The bare results are shown in \cref{tbl:fp:results}. 

\begin{figure}[tb]
  \centering
  \begin{subfigure}[t]{.13 \linewidth} 
    \centering
    \includegraphics[page=3]{c4-f3-opt-solution}
  \end{subfigure}
  \begin{subfigure}[t]{.41 \linewidth} 
    \centering
    \includegraphics[page=1]{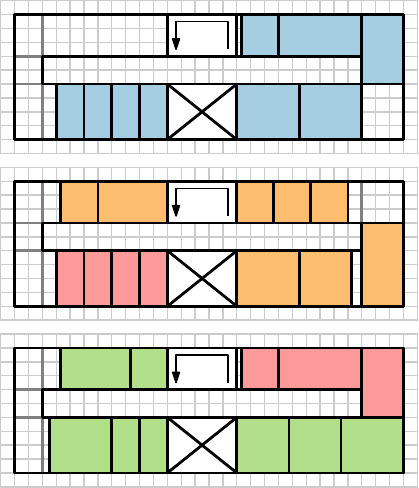}
    \caption{Optimal solution by \FPILP.}
    \label{fig:solution:sM-3M:ilp}
  \end{subfigure}
  \hfill
  \begin{subfigure}[t]{.41 \linewidth} 
    \centering
    \includegraphics[page=2]{c4-f3-opt-solution}
    \caption{Manually post-processed solution.}
    \label{fig:solution:sM-3M:manual}
  \end{subfigure}
    \caption{The solutions found by \FPILP, 
    like the optimal solution for the problem instance \sMM here,
    should not be interpreted as final floor plans.}
    \label{fig:solution:sM-3M}
\end{figure}

\FPILP found solutions for small to medium-sized instances but it
managed to solve only the smallest instance \sMM to optimality.
The resulting floor plans, found within five minutes, 
are shown in \cref{fig:solution:sM-3M:ilp}.
Knowing that at least one group needs to be split to two floors,
the results looks satisfactory.

\begin{figure}[tb]
	\centering
	\begin{subfigure}[t]{.1 \linewidth} 
    	\centering
    	\includegraphics[page=7,width=0.6\textwidth]{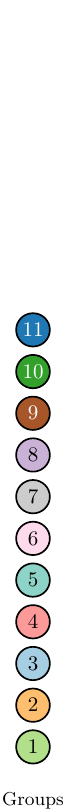}
    \end{subfigure}
	\begin{subfigure}[t]{.26 \linewidth}
		\centering
		\includegraphics[page=2,width=0.9\textwidth]{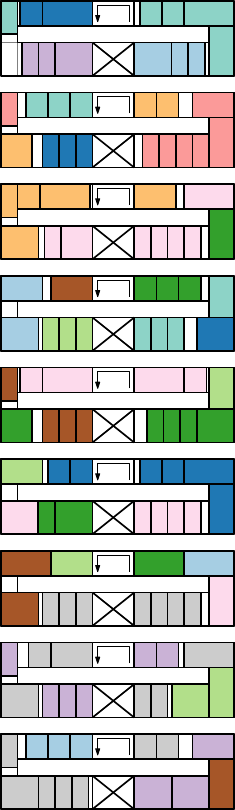}
		\caption{\FPILP (\SI{1}{\hour}): 13816}
		\label{fig:solution:c11-f9-ilp:1h}
	\end{subfigure}
	\hfill
	\begin{subfigure}[t]{.26 \linewidth}
		\centering
		\includegraphics[page=3,width=0.9\textwidth]{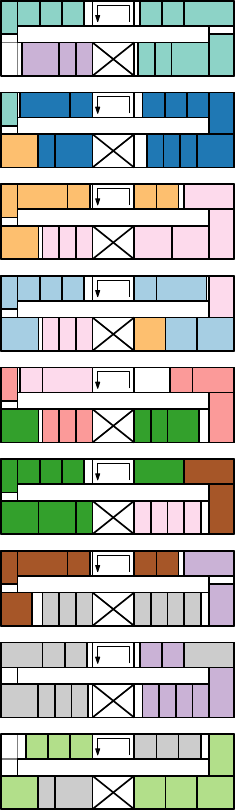}
		\caption{\FPILP (\SI{12}{\hour}): 6754}
		\label{fig:solution:c11-f9-ilp:12h}
	\end{subfigure}
	\hfill
	\begin{subfigure}[t]{.26 \linewidth}
		\centering
		\includegraphics[page=8,width=0.9\textwidth]{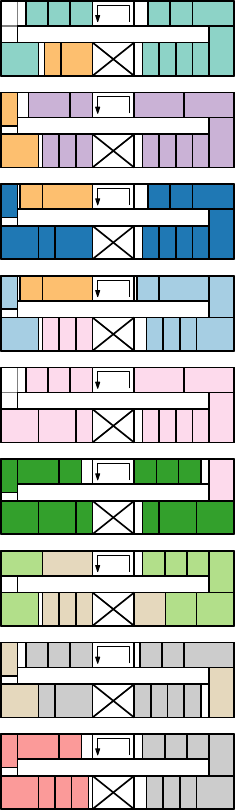} 
		\caption{\FPILP (\SI{66}{\hour}): 4235}
		\label{fig:solution:c11-f9-ilp:66h}
	\end{subfigure}
	\caption{Solutions for \MM computed by \FPILP with different time limits.}
	\label{fig:solution:c11-f9-ilp}
\end{figure}

Concerning the medium-sized instances (\textsf{M-XX}), 
\FPILP performed significantly better with the higher time limit.
For example, for the instance {\MM}, the floor plans 
after \SI{1}{\hour}, \SI{12}{\hour}, and \SI{66}{\hour} of computation
time are shown in \cref{fig:solution:c11-f9-ilp}.
We observe that the floor plans after \SI{12}{\hour} and especially after \SI{66}{\hour} 
appear considerably tidier than after \SI{1}{\hour},
though still not fully satisfactory.

The situation is worse for larger problem instances.
Namely, for \MCL, no solution was found even after \SI{12}{\hour}.
For \CL, \FPILP found solutions but ran out of memory after roughly \SI{8}{\hour}. 
The RAM usage shown in \cref{tbl:fp:ram} suggests that this is a systematic problem.
Hence, these results indicate that it is necessary to break down
large problem instances into smaller ones
by computing a floor assignment first.

\begin{table}[tb]
  \centering
  \setlength{\tabcolsep}{1ex}
\begin{tabular}{@{}lrrrrrr@{}}
\toprule
Instance 				& \sMM             & \MXL & \MM        & \MS         & \CL        & \MCL\\
\midrule
\FPILP (\SI{12}{\hour})	& 			  	   & 848  & 6754 	   & 3398 		 & \text{\color{BrickRed} ooMem}& -- 	\\
\FPILP (\SI{1}{\hour}) 	& \textbf{241} 	   & 1516 & 13816 	   & 3877		 & 85142 	   & -- 	 \\
\FAILP (\SI{1}{\hour}) + \FPILP & \bad{458}& 642  & \bad{5207} & \bad{5090}	 & \bad{19934} & \bad{15737}  \\
\FAHEU + \FPILP 		& 512       	   & 1398 & \bad{6436} & \bad{7538}	 & \bad{23195} & \bad{26300}	 \\
\bottomrule
\end{tabular}
	\medskip
	\caption{Best solutions found for different approaches and instances.
    The two versions of \FPILP differ only in their time limit.  
    Results written in \bad{red italic} required scaling of room sizes to admit solutions.}
\label{tbl:fp:results}
\end{table}
\begin{table}[tb]
  \centering
  \setlength{\tabcolsep}{1ex}
\begin{tabular}{@{}lrrrrrr@{}}
\toprule
Instance 			   & \sMM  & \MXL   & \MM   & \MS  & \CL      & \MCL\\
\midrule
\FPILP (\SI{12}{\hour})& -- 	   & 900 MB & 16 GB & 3 GB & $>79$ GB & 66 GB\\
\FPILP (\SI{1}{\hour}) & 10 MB & 900 MB & 15 GB & 3 GB & 13 GB 	  & 6 GB\\
\bottomrule
  \end{tabular}
  \medskip
  \caption{Maximum amount of memory used by \FPILP (excluding overhead).}
  \label{tbl:fp:ram}
\end{table}

\paragraph{Floor assignment as a preprocessing step.}
The two floor assignment algorithms \FAILP and \FAHEU distribute
rooms to floors such that \FPILP has to solve only single-floor instances.
We compare the solutions found by either combination 
with those found by \FPILP alone. 
To this end, we combined the solutions for~the single-floor instances
to a solution of the original multi-floor instance
and computed the quality according to the objective function of~\FPILP; 
see \cref{tbl:fp:results}.
\FAILP instances were again solved with CPLEX and DFS-like settings.

Both floor assignment approaches
made it possible to find solutions for the two largest problem instances
where \FPILP on its own did not succeed. 
Furthermore, the combination of \FAILP and \FPILP 
found a better solution to \MXL than \FPILP alone
and this in a total of only 14 minutes.
In general, the combination of \FAILP and \FPILP achieved 
better results than the combination of \FAHEU and \FPILP.
However, there is a caveat concerning these results.
For most problem instances, it was necessary to downscale 
the room sizes for several floors such that they would admit solutions.
In this regard, \FAHEU performed better than \FAILP;
we discuss this in more detail below.

\begin{figure}[tb]
	\centering
	\begin{subfigure}[t]{.1 \linewidth} 
    	\centering
    	\includegraphics[page=7,width=0.6\textwidth]{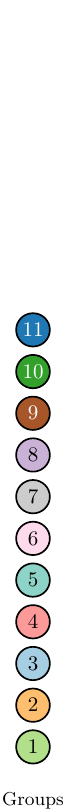}
    \end{subfigure}
	\begin{subfigure}[t]{.26 \linewidth}
		\centering
		\includegraphics[page=4,width=0.9\textwidth]{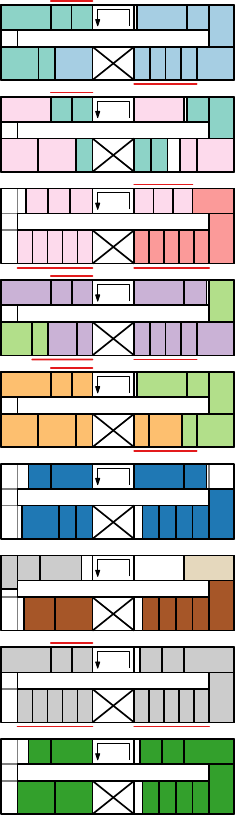}
		\caption{\FAILP + \FPILP: \bad{5207}}
		\label{fig:solution:c11-f9-fa:ilp}
	\end{subfigure}
	\hfill
	\begin{subfigure}[t]{.26 \linewidth}
		\centering
		\includegraphics[page=5,width=0.9\textwidth]{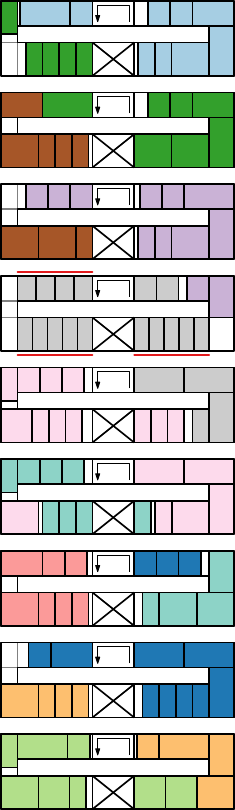}
		\caption{\FAHEU + \FPILP: \bad{6436}}
		\label{fig:solution:c11-f9-fa:heu}
	\end{subfigure}
	\hfill
	\begin{subfigure}[t]{.26 \linewidth}
		\centering
		\includegraphics[page=6,width=0.9\textwidth]{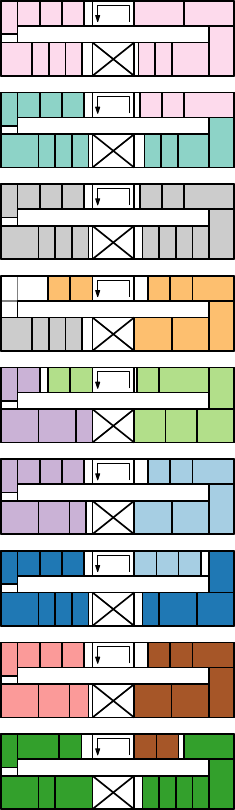}
		\caption{Manual: 5144}
		\label{fig:solution:c11-f9-fa:manual}
	\end{subfigure}
	\caption{Solutions of \MM found by first computing a floor assignment 
	with \FAILP and \FPILP as well as a manually crafted solution.
	Red line segments indicate that the sizes of the rooms next to
        the segments had to be scaled down slightly.}
	\label{fig:solution:c11-f9-fa}
\end{figure}

\Cref{fig:solution:c11-f9-fa:ilp,fig:solution:c11-f9-fa:heu}
show the floor plans computed for \MM
with the help of \FAILP and \FAHEU, respectively.
We find that both solutions appear tidier
than those found by \FPILP alone (\cref{fig:solution:c11-f9-ilp}).
Moreover, since each group appears on at most two floors,
these solutions appear closer to what a human planner would construct.
We further want to point out that, for \FAHEU,
only one floor required downscaling of room sizes.

Next, we consider the runtime of the floor assignment approaches.
While the runtime of \FAHEU is negligible,
we gave \FAILP a time limit of one hour.
\Cref{tbl:fa:runtimes} shows the time required by \FAILP
to find optimal solutions and to prove that they are optimal.
Often, \FAILP found the optimal floor assignment solution 
within a few seconds for all instances,
but required a few minutes to prove their optimality;
for \MS it was not able to prove optimality even after three days.

For both \FAHEU and \FAILP as preprocessing steps,
\cref{tbl:fa:runtimes} shows the runtime
required by \FPILP to find and prove optimal solutions (aggregated over all floors).
\FPILP solved most buildings in less than \SI{10}{\second}.
Only the instances with the large empty floor
plan~$f_\mathrm{XL}$ still took a few minutes to be solved.
Nevertheless, we can conclude that first computing a floor assignment
and then using \FPILP for single floors
yields a tremendous speed~up.

\begin{table}[tb]
  \centering
  \setlength{\tabcolsep}{1ex}
\begin{tabular}{@{}lrrrrrr@{}}
\toprule
Instance 			& \sMM & \MXL   & \MM       & \MS   & \CL   & \MCL\\
\midrule
\FAILP finding OPT 	& 0.01 & 0.35 	&    1.09 	& 5.47 	&  2.36 & 4.70\\
\FAILP proving OPT	& 0.18 & 0.35 	& 2134.32	& -- 	& 34.84 & 1067.23\\
\FPILP (\FAILP) 	& 1.33 & 808.69 &    3.10 	& 0.37	&  2.32 & 8.97\\
\FPILP (\FAHEU) 	& 1.22 & 170.44 &    4.06 	& 0.63 	&  4.27 & 10.71 \\
\bottomrule
  \end{tabular}
  \medskip
  \caption{Runtimes of \FAILP in seconds for finding and for proving
    solutions optimal (rows~1 and 2); aggregated runtimes for the \FPILP
    runs over all floors (rows~3 and~4).}
  \label{tbl:fa:runtimes}
\end{table}

\paragraph{Scaling to the rescue.}
Since the floor assignment approaches do not take the empty floor plans 
into account when distributing the rooms it may happen 
that a floor assignment solution results in unsolvable floor planning instances.
For such an unsolvable instance, we scaled down all room sizes with the same scaling factor 
such that the instance became solvable.
For the different problem instances, 
\cref{tbl:floorassignment:scaling} shows the worst needed scaling factor 
among all floors and the number of floors that required scaling.

Taking a closer look at the solutions in \cref{fig:solution:c11-f9-fa},
we observe that downscaling was not only necessary because of a lack of empty space, 
but also because of a bad distribution of room sizes. 
For example, the sixth floor in the \FAHEU solution has many small rooms but only one large room
and therefore not all corners can be occupied.
The same problem occurred on the second floor in the \FAILP solution.
We further find the \FAILP solutions mostly required downscaling because of tightly packed floors.
By design, this is less likely to happen with \FAHEU.

\begin{table}[tb]
  \centering
  \setlength{\tabcolsep}{1ex}
\begin{tabular}{@{}llrrrrrr@{}}
\toprule
Instance 		&& \sMM & \MXL   & \MM       & \MS   & \CL   & \MCL\\
\midrule
	\parbox[t]{12mm}{\multirow{2}{*}{\FAILP}} 
        & Scaling factor & 165/171 	& -- 		& 153/171 	& 82/99 	& 207/311 	& 207/311 	\\
	  & \# floors scaled & 1 	 	& 0 		& 6 		& 9 		& 6 		& 7 		\\\hline
	\parbox[t]{12mm}{\multirow{2}{*}{\FAHEU}} 
        & Scaling factor & -- 		& -- 		& 144/171 	& 94/99		& 207/311 	& 207/311	\\
	  & \# floors scaled & 0 		& 0 		& 1 		& 1			& 3 		& 8 		\\
\bottomrule
  \end{tabular}
  \medskip
  \caption{Worst needed scaling factors and the number of floors that needed scaling
  when solving the (single-floor) \textsc{Floor Planning} instances created 
  by floor assignment solutions.}
  \label{tbl:floorassignment:scaling}
\end{table}

\paragraph{Algorithms vs.\ human.}
For the problem instance \MM
we also constructed a floor plan manually; see \cref{fig:solution:c11-f9-fa:manual}.
Compared to the floor plans computed by the three different algorithmic approaches
(see again \cref{fig:solution:c11-f9-ilp,fig:solution:c11-f9-fa}),
the manual floor plan appears more coherent in the following sense.
Groups form only few clusters and do not alternate in the cyclic order
around the corridors (whereas they do in the ground-floor of
\cref{fig:solution:c11-f9-ilp:12h}).
Interestingly, the floor plan computed by \FPILP after \SI{66}{\hour}
(\cref{fig:solution:c11-f9-ilp:66h}) achieved a lower score (4235)
than the objective function value of the manual solution (5144),
but appears less coherent.  Note, however, that \FPILP placed more
groups within one floor (7 vs.\ 6).  On the other hand, in
\cref{fig:solution:c11-f9-ilp:66h}, the small orange group~2 is spread
over four floors (with 1, 1, 2, 1 objects).  This yields~14 times the
intra-floor distance, that is, the distance between two consecutive
floors (which we set to \SI{20}{\meter}).  Although the larger violet group~8 in
\cref{fig:solution:c11-f9-fa:manual} is spread over just two floors,
where it occupies four objects on each floor.
This yields more, namely 16 times the intra-floor distance.  If we want to
favor solutions that avoid spreading groups over more than two floors,
we simply have to replace our linear floor distance function
(see p.~\pageref{page:distance}) by, say, a quadratic one
(e.g., $\abs{j-i}^2\cdot\SI{20}{\meter}$ for the distance from
floor~$i$ to floor~$j$). 

\paragraph{Post-processing of ILP solutions.}
A solution computed with \FPILP only specifies a room placement,
which, by the definition in \cref{sec:model}, 
is only a mapping of rooms to edges and corners
that does not include an ordering of rooms along the same edge.
For the floor plans depicted in \cref{fig:solution:sM-3M,fig:solution:c11-f9-ilp,fig:solution:c11-f9-fa},
we ordered rooms along the same edge manually in a post-processing step.
In our experiments, we observed that in most cases only one group, rarely two groups,
and only once three groups were present at the same edge.
We hence expect the task of ordering rooms along an edge to be trivial in practice.

\section{Concluding Remarks} 
\label{sec:conclusion}

In this paper, we introduced the \textsc{Floor Planning with Group Proximity} problem
and described an ILP formulation to solve it.
We tested our ILP on realistic test data using the ILP solver CPLEX.
Our experiments showed that small problem instances (28 rooms, 4 groups) 
can be solved to optimality within minutes and with satisfactory results.
While medium-sized instances (122 rooms, 11 groups) can still be solved within a few hours,
the complexity of the problem makes it unfeasible to tackle larger instances directly.

We further showed that large (multi-floor) instances of the floor planning problem
can be tackled by first solving the respective \textsc{Floor Assignment with Group Proximity} problem
and then solving the resulting (single-floor) instances as before.
We have formulated the floor assignment problem, too, as an ILP,
but we also suggested a greedy heuristic for it.
Our experiments demonstrated that splitting instances via a floor assignment
wasn't only much faster, but also improved some of the results for
medium-sized instances that \FPILP computed within a fixed runtime limit.

Let us consider the optimal solution found for the smallest instance \sMM 
by \FPILP in \cref{fig:solution:sM-3M:ilp}.
Since the ILP only optimizes according to an objective function,
we should not be surprised to find, for example, gaps between rooms or
corner rooms that cannot be entered.
A planner would directly avoid these and other deficiencies.
However, such deficiencies can easily be eradicated in a (manual) post-processing step
as shown by the floor plan in \cref{fig:solution:sM-3M:manual}.
As mentioned in the introduction, if a user is not satisfied with
an optimal solution (say, to an ILP), then this is not a failure of the solver, 
but a failure of the model and the corresponding objective function.
Here, we decided to keep the model as simple as possible to test
the potential computational limitations of our approach.

The proposed variants of the floor planning problem and our
algorithmic approaches have demonstrated their potential to support planners 
in the design of architectural floor plans of office buildings.  
We stress that formulating the floor planning problem as an ILP has
the advantage that a planner may add further requirements. 
For example, one can easily add constraints that force 
two given rooms to be adjacent, extend the model to less constrained
building outlines, or allow rooms to occupy two corners.

\pdfbookmark[1]{References}{References} 
\bibliographystyle{splncs04}
\bibliography{sources}

\end{document}